\documentclass[9pt,twocolumn,twoside]{osajnl}
\journal{ol} % Choose journal (ao,jocn,josaa,josab,ol,optica,pr)

\setboolean{shortarticle}{true}
\usepackage{lineno}
\title{Frozen spatial coherence}

\author[1]{M. A. Pinto}
\author[1,*]{P. A. Brandão}
\affil{Instituto de F\'isica, Universidade Federal de Alagoas, Macei\'o, 57072-900, Brazil.}
\affil[*]{Corresponding author: paulo.brandao@fis.ufal.br}

\begin{abstract}
Inspired by the concept of coherent frozen waves, this paper introduces one possible theoretical framework of its partially coherent version, a frozen spatial coherence, in which a desired two-point correlation structure of an optical field is created on the propagation axis by superposing partially coherent zero-order Bessel beams. It is shown that the cross-spectral density can be given a description in terms of a two-dimensional Fourier series, analogous to the one-dimensional approach of coherent frozen waves. The formalism is applied to the design of a partially coherent field which is highly coherent only if the pair of points in the propagation axis belong to a predetermined and finite range and highly incoherent outside that range.
\end{abstract}

\setboolean{displaycopyright}{true}

\begin{document}

\maketitle

%\section{Introduction}

The generation and manipulation of light beams in a controllable way is one of the most important goals in optics. A big step in this direction was taken by J. Durnin with his discovery of nondiffracting waves, in particular, the Bessel beam solution \cite{durnin1987exact,durnin1987diffraction}. By controlling the diffractive properties of propagating waves in free-space, Durnin demonstrated that an optical field, with a variable transverse intensity profile, can travel a long distance without significant changes in its intensity when compared to the propagation of a Gaussian beam. The stability of intensity during propagation is reflected in the phenomenal ability of the beam to self-reconstruct when obstructed by opaque obstacles in its propagation path \cite{bouchal1998self}. This self-reconstructing property was used profitably in microscopy applications \cite{fahrbach2010microscopy}. 

% Parte removida sugerida pelo Referee
%A more general framework in achieving unprecedented control over the interaction between electromagnetic fields and matter has been developed and given the name \textit{transformation optics} \cite{LEONHARDT200969,mccall2018roadmap}. In this context, it was possible to suggest and implement interesting optical effects such as perfect invisibility devices \cite{schurig2006metamaterial} and perfect lenses based on metamaterials \cite{pendry2000negative}, to cite a couple of examples.

A less dramatic but still interesting approach to the control of the intensity of an electromagnetic beam has been given by the concept of frozen waves \cite{zamboni2005theory}. The authors demonstrated that by performing a coherent superposition of zero-order Bessel beams, one could obtain a desired, predetermined, intensity pattern profile on the axis of the propagation direction. With the use of a spatial light modulator, the theory has been verified in a laboratory \cite{vieira2012frozen}. An obvious application of these ideas is in the area of optical trapping where the control of on-axis intensity plays a major role. The trapping of microparticles using frozen waves has been demonstrated experimentally in recent years \cite{suarez2020experimental}. The authors recognized the possibility of a stable optical trap and transverse guidance. By generalizing the theory to include superposition of Bessel beams of arbitrary order, it was demonstrated that the topological charge of the overall beam can be controlled during propagation, thus providing a more systematic way to tune its orbital angular momentum properties \cite{dorrah2016controlling}.

In the case of frozen waves, it was assumed that the time behavior of the electromagnetic field is deterministic in order to derive the main results of the theory. In other words, the resulting beam is completely coherent. It is well known that in realistic physical situations, the time evolution of the field (in one or more positions) is described by a random process \cite{mandel1995optical,goodman2015statistical,wolf2007introduction}. The classical theory of optical coherence is concerned with the characterization of correlations between various points in the field. A very readable account of the main applications of optical coherence has been published recently \cite{korotkova2020applications}. It is thus only natural to ask if the concepts lying behind the theory of frozen waves can be incorporated (in some sense) into the partially coherent domain. The objective of this paper is to show that a new kind of control over the coherence properties of a propagating partially coherent beam can be achieved by performing superpositions of partially coherent Bessel beams.

Below we formulate the theory in which a desired, predetermined, cross-spectral density profile $W(\mathbf{r}_1,\mathbf{r}_2,\omega)$ can be constructed through the incoherent superposition of Bessel beams. The analysis closely follows the one developed for coherent frozen waves. In particular, we demonstrate that the cross-spectral density function can be given a representation in which it exactly matches a two-dimensional Fourier series. Next, the formalism is applied to a specific example consisting of a partially coherent field in which there are strong correlations between a pair of points on the beam axis only if these points are inside some fixed region. These statements will be formally described below.

%\section{General theory}
We start with the general representation for the cross-spectral density $W(\mathbf{r}_1,\mathbf{r}_2,\omega)$ for partially coherent beams in terms of Bessel functions \cite{kowarz1995bessel},
\begin{multline}\label{w1}
    %\begin{split}
        W(\mathbf{r}_1,\mathbf{r}_2) = \sum_{p,q = -\infty}^{\infty} \int_0^{\infty}d\alpha_1\int_0^{\infty}d\alpha_2 \\
        \times  C_{pq}(\alpha_1,\alpha_2) B_p^*(\mathbf{r}_1;\alpha_1)B_q(\mathbf{r}_2;\alpha_2),
    %\end{split}
\end{multline}
where $B_s(\mathbf{r};\alpha) = J_s(k\alpha\rho)\exp(is\phi)\exp(ik\beta z)$ represents the coherent Bessel beam solution of the Helmholtz equation with
\begin{equation}\label{betacond}
\beta = \begin{cases}
\sqrt{1 - \alpha^2} &\text{if $\alpha \leq 1$}\\
i\sqrt{\alpha^2 - 1} &\text{if $\alpha > 1$}
\end{cases}
\end{equation}
and $(\rho,\theta,z)$ denotes a point in space in cylindrical coordinates. The dependence of the cross-spectral density on the frequency $\omega$ is omitted for brevity in what follows. Depending on the functional form of $C_{pq}(\alpha_1,\alpha_2)$, \eqref{w1} is able to represent a large class of coherent and partially coherent wavefields. For example, in the special case where $C_{pq}(\alpha_1,\alpha_2) = c_p(\alpha_1)c_q(\alpha_2)$ is separable, the cross-spectral density represents a complete coherent beam with field amplitude $\psi(\mathbf{r})$ of the form \cite{kowarz1995bessel}
\begin{equation}
    \psi(\mathbf{r}) = \sum_{p = -\infty}^{\infty} \int_0^{\infty} c_p(\alpha) B_p(\mathbf{r};\alpha)d\alpha,
\end{equation}
the special case $c_p(\alpha) = \delta_{pm}\delta(\alpha - \alpha_0)$ being the familiar Bessel beam of order $m$ and transverse wavevector $k\alpha_0$. In general, for a nonseparable functional form of $C_{pq}(\alpha_1,\alpha_2)$ a partially coherent field is obtained. In terms of genuine correlation functions described in \cite{gori2007devising}, \eqref{w1} can be obtained from the more general structure
\begin{equation}\label{wgenuine}
    W(\pmb{r}_1,\pmb{r}_2) = \int p(v)H^*(\pmb{r}_1,v)H(\pmb{r}_2,v)dv
\end{equation}
by choosing $H(\pmb{r},v) = \sum_{p=-\infty}^{\infty} \int_0^{\infty} C_p(\alpha,v) B_p(\mathbf{r};\alpha)$ and noting that $C_{pq}(\alpha_1,\alpha_2) = \int p(v)C_p^*(\alpha_1,v)C_q(\alpha_2,v)dv$ with $p(v) > 0$. Correlation profiles involving Bessel functions have also been considered in \cite{ponomarenko2007dark,zhu2019experimental} where the authors demonstrate dark and antidark diffraction-free beams which are represented by $z$-independent cross-spectral densities.

Since our main interest here is in the correlations existent along the $z$ axis between two points $z_1$ and $z_2$ with $\rho_1 = \rho_2 = 0$, a direct consequence of   \eqref{w1} is that
\begin{multline}\label{wz}
    %\begin{split}
        W(z_1,z_2) = \int_0^{\infty}d\alpha_1\int_0^{\infty}d\alpha_2 C_{00}(\alpha_1,\alpha_2) \\
        \times \exp[-ik(\beta_1 z_1 - \beta_2 z_2)],
    %\end{split}
\end{multline}
so that only the coefficient $C_{00}(\alpha_1,\alpha_2)$ contributes to the correlations along the $z$ axis. To obtain \eqref{wz} we used the fact that $J_s(0) \sim \delta_{s0}$. This result suggests an approach reminiscent of the so-called frozen waves \cite{zamboni2005theory} if we assume $C_{pq}$ to be nonzero only for $p = q = 0$ and consider
\begin{equation}\label{c00}
    C_{00}(\alpha_1,\alpha_2) = \sum_{n,m = -\infty}^{\infty} d_{nm}\delta(\alpha_1 - \tilde{\alpha}_n)\delta(\alpha_2 - \tilde{\alpha}_m),
\end{equation}
where $\{ d_{nm} \} $ is a set of coherence parameters and $\{ \tilde{\alpha}_p \} $ is a discrete set of transverse wavevectors. After substituting \eqref{c00} into \eqref{wz} we obtain
\begin{equation}\label{wfourier}
    W(z_1,z_2) = \sum_{n,m = -\infty}^{\infty} d_{nm}\exp[-ik(\tilde{\beta}_n z_1 - \tilde{\beta}_m z_2)],
\end{equation}
which can be written in the exact form of a Fourier series in two variables if we choose $k\tilde{\beta}_l =  2\pi l/L$, where $L$ is the length of interest in the $z$ axis with $z_1,z_2 \in [0,L]$. However, this is not desirable since we are interested in cases where $\tilde{\beta}_l$ is positive [and $\tilde{\alpha}_l < 1$ so that evanescent waves are neglected, see   \eqref{betacond}]. We thus follow \cite{zamboni2005theory} and consider 
\begin{equation}\label{betacond2}
    k\tilde{\beta}_l = Q + \frac{2\pi l}{L},
\end{equation}
where $Q$ is a fixed parameter chosen to guarantee that $\tilde{\beta}_l > 0$, or, $Q > 2\pi l_{\text{max}}/L$ where $l_{\text{max}}$ is the maximum value of $l$ in the computational summation. The condition $\tilde{\beta}_l < 1$ must also be satisfied. By substituting \eqref{betacond2} into \eqref{wfourier} we obtain
\begin{equation}\label{eq1}
    \begin{split}
        \exp[iQ(z_1 &-  z_2)]W(z_1,z_2) \\
        &= \sum_{n,m=-\infty}^{\infty}d_{nm} \exp\left[ -\frac{2\pi i}{L}(nz_1 - mz_2)  \right].
    \end{split}
\end{equation}
The function on the left-hand side of   \eqref{eq1} is thus represented as a Fourier series. Once $W(z_1,z_2)$ is given, the coherence parameters $d_{nm}$ can be calculated by
\begin{equation}\label{dnm}
\begin{split}
    d_{nm} &= \frac{1}{L^2}\int_0^L dz_1 \int_0^L dz_2 \exp[iQ(z_1 -  z_2)] \\
    &\times W(z_1,z_2)\exp\Bigg[\frac{2\pi i}{L}(nz_1 - mz_2)\Bigg].
\end{split}
\end{equation}
Equation \eqref{dnm} can be used to obtain the coherence parameters $d_{nm}$ if a desired, predetermined, correlation profile $W(z_1,z_2)$ is given.

In closing this discussion, we remark that the coherence parameters $d_{nm}$ cannot be chosen arbitrarily, i.e., they must satisfy certain symmetry requirements. Since the spectral density $S(\mathbf{r}) = W(\mathbf{r},\mathbf{r})$ is real and positive, substituting $z_1 = z_2 = z$ in   \eqref{eq1} gives the condition $d_{nm} = d_{mn}^*$. Then,   \eqref{dnm} satisfies this constraint if $W(z_1,z_2)$ is symmetric with respect to the exchange between $z_1$ and $z_2$: $W(z_1,z_2) = W(z_2,z_1)$. We deal only with symmetric correlation functions in what follows.

%\section{Application and discussion}

\begin{figure}[hbt]
    \centering
    \includegraphics[scale=0.17]{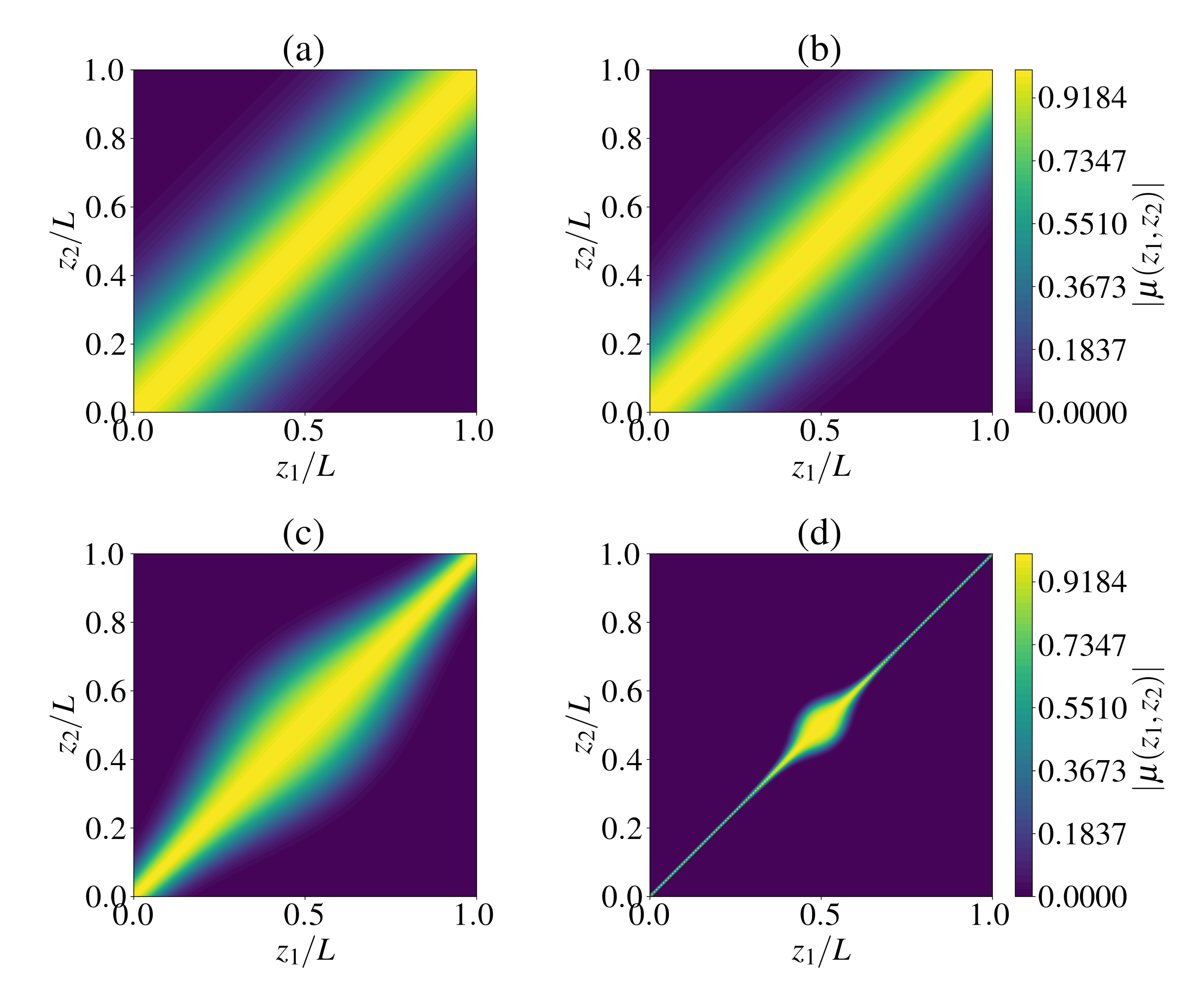}
    \caption{Plots of the desired spectral degree of coherence given by \eqref{mu1} for several values of $\kappa$. Parameters: $\Delta = 50$, $L = 80\pi$, (a) $\kappa=0$, (b) $\kappa=10^{-5}$, (c) $\kappa=10^{-4}$, (d) $\kappa=10^{-2}$. } 
    \label{fig1}
\end{figure}

We are now ready to discuss an example in which a specific correlation profile [  \eqref{mu1} below] is obtained by the superposition of partially coherent Bessel beams. Suppose we wish to generate a partially coherent wave propagating in the $z$ direction such that it is highly coherent only around some region on the $z$ axis, say, $z \sim L/2$, where $z \in [0,L]$ is the propagation range considered. By this we mean that the optical field at any chosen pair of points $(z_1,z_2)$ is highly coherent only if $z_1 \sim L/2$ and $z_2 \sim L/2$. To make things more quantitative, it is convenient to deal with the spectral degree of coherence $\mu(\mathbf{r}_1,\mathbf{r}_2)$, defined by
\begin{equation}\label{defmu}
    \mu(\mathbf{r}_1,\mathbf{r}_2) = \frac{W(\mathbf{r}_1,\mathbf{r}_2)}{\sqrt{S(\mathbf{r}_1)}\sqrt{S(\mathbf{r}_2)}},
\end{equation}
where $|\mu(\mathbf{r}_1,\mathbf{r}_2)| \leq 1$ is always satisfied for arbitrary points in the optical field. This quantity is defined in the space-frequency formalism of second-order classical coherence and is thus valid for a fixed frequency $\omega$. Since we are mainly interested in the correlations existing in the optical field, we choose $S(z) \sim S_0$ on the $z$ axis, with $S_0$ a positive constant, and define the function 
\begin{equation}\label{mu1}
    \mu(z_1,z_2) = \exp\left[ -\frac{(z_1 - z_2)^2}{\sigma^2(z_1,z_2)} \right],
\end{equation}
where
\begin{equation}\label{eqsigma}
    \sigma(z_1,z_2) = \frac{\sqrt{2}\Delta}{1 + \kappa\left[\left( z_1 - \frac{L}{2} \right)^2 + \left( z_2 - \frac{L}{2} \right)^2\right] }.
\end{equation}
The suggested model represented by \eqref{eqsigma} expresses the width $\sigma(z_1,z_2)$ of the correlations in the $(z_1,z_2)$ plane where $\kappa$ is a real parameter. The special case $\kappa = 0$ represents the Gaussian-Schell model with constant width $\Delta$. Note that \eqref{mu1} satisfies $\mu(z_1,z_2) = \mu(z_2,z_1)$ and the physical condition $\mu(z,z) = 1$, as required. Several two-dimensional plots of \eqref{mu1} are shown in Figure \ref{fig1} for increasing values of $\kappa$. The symmetry requirements on $\mu$ can be easily seen in these plots.

Once the coherence parameters $d_{nm}$ are determined by the use of   \eqref{dnm}, we are able to calculate the approximate cross-spectral density $W_{\text{app}}(\mathbf{r}_1,\mathbf{r}_2)$, given by
\begin{equation}\label{wapp}
    \begin{split}
        W_{\text{app}}(\mathbf{r}_1,\mathbf{r}_2) &= \sum_{n,m = -N}^{N}d_{nm}J_0(k\tilde{\alpha}_n\rho_1)J_0(k\tilde{\alpha}_m\rho_2)\\
        &\times \exp[-ik(\tilde{\beta}_n z_1 - \tilde{\beta}_m z_2)],
    \end{split}
\end{equation}
where $N$ $(-N)$ is the maximum (minimum) value of $n$ and $m$. The cross-spectral density given by \eqref{wapp} can also be obtained directly from \eqref{wgenuine} by selecting $H(\pmb{r},v) = \sum_{n=-N}^N D_n(v)J_0(k\alpha _n\rho)e^{ik\beta_n z }$ where $d_{nm} = \int p(v)D_n^*(v)D_m(v)dv$ such that $d_{nm}^* = d_{mn}$. Therefore, there are $(2N + 1)^2$ coherence parameters $d_{nm}$ composing our approximation to the cross-spectral density. The approximate spectral density $S_{\text{app}}(\mathbf{r})$ is readily evaluated as
\begin{equation}\label{Sapp}
    \begin{split}
        S_{\text{app}}(\mathbf{r}) = &\sum_{n,m = -N}^{N}  d_{nm}J_0(k\tilde{\alpha}_n\rho)J_0(k\tilde{\alpha}_m\rho)\\
        &\times \exp[-ikz(\tilde{\beta}_n - \tilde{\beta}_m)].
    \end{split}
\end{equation}

\begin{figure}[hbt]
    \centering
    \includegraphics[scale=0.23]{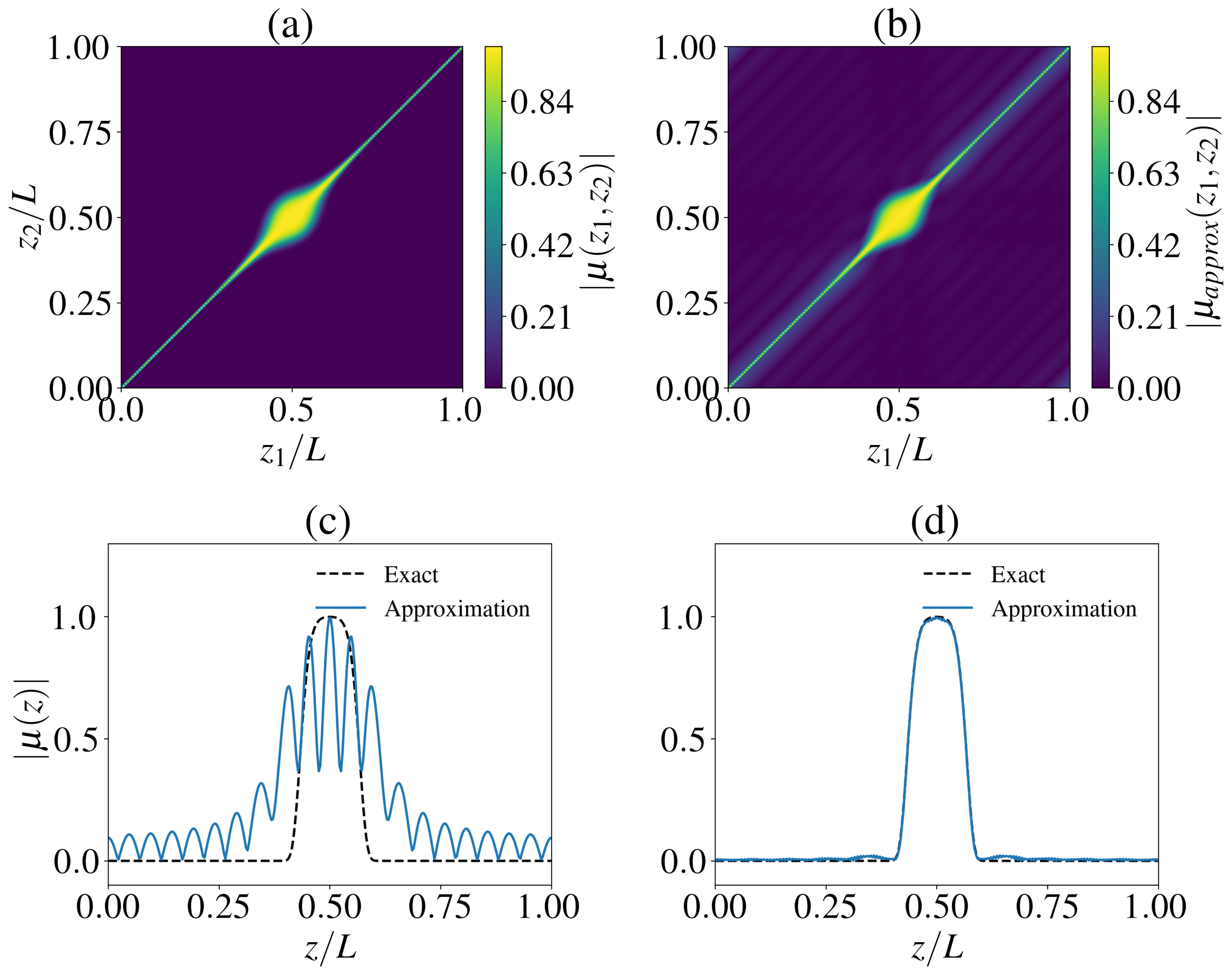}
    \caption{(a) Plot of the desired correlation profile $|\mu(z_1,z_2)|$ given by \eqref{mu1} and (b) its approximation $|\mu_{\text{app}}(z_1,z_2)|$ with $N = 80$. The plots in the bottom part of the figure show the comparison for a slice at $z_2 = L/2$ for (c) $N=10$ and (d) $N=80$. Parameters used: $L = 80\pi$, $Q = 1/2 + 2\pi N/L$, $\kappa = 10^{-2}$, $\Delta = 50$.} 
    \label{fig2}
\end{figure}

\begin{figure}[hbt]
    \centering
    \includegraphics[scale=0.4]{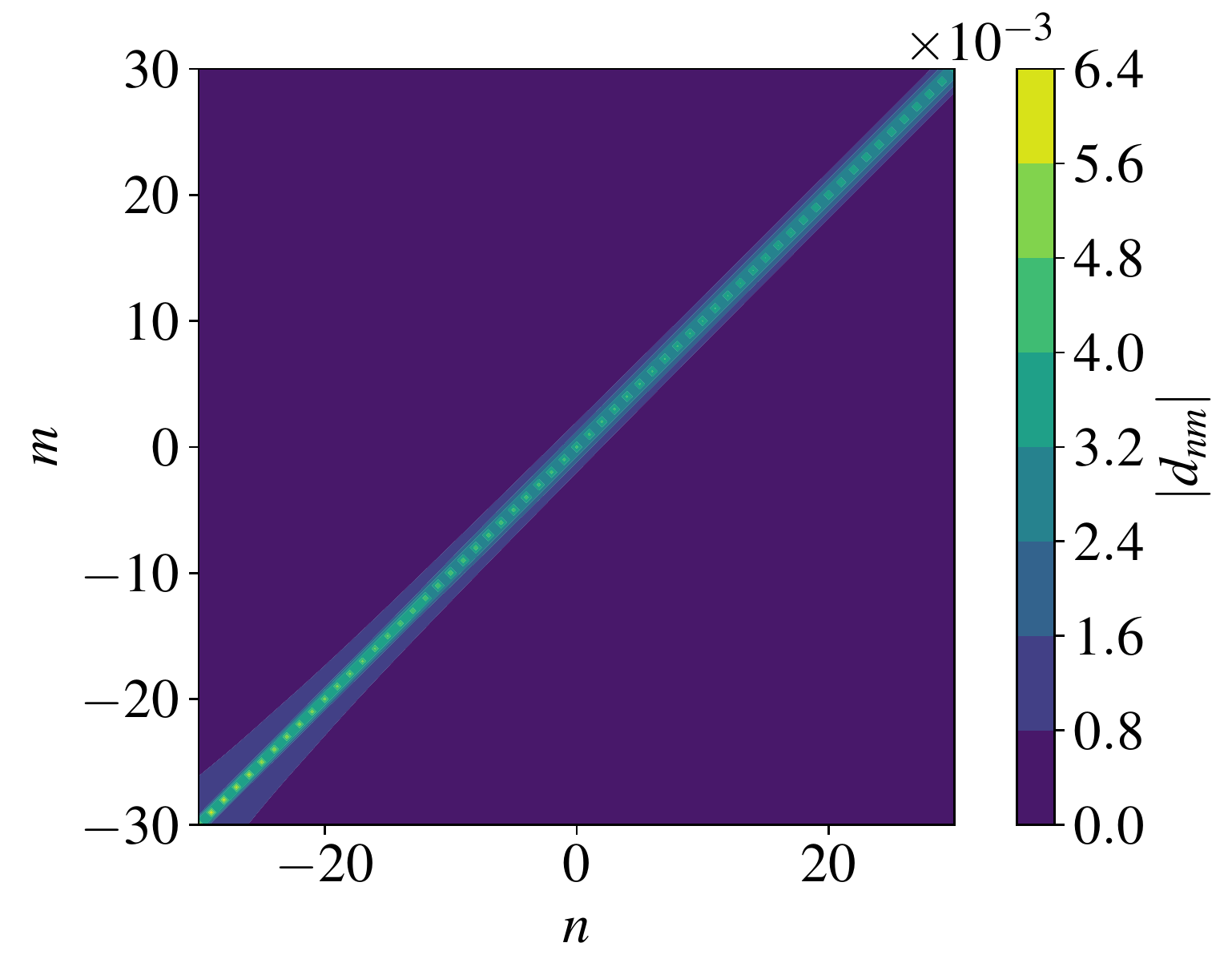}
    \caption{Graphical illustration of the coherence parameters $|d_{nm}|$ as a function of $n$ and $m$. Parameters: $\Delta = 50$, $L = 80\pi$, $\kappa=10^{-2}$, $N=30$, $Q = 1/2 + 2\pi N / L$.} 
    \label{fig3}
\end{figure}

Part (a) of Figure \ref{fig2} shows the the desired profile obtained from   \eqref{mu1} with $\kappa = 10^{-2}$, $\Delta = 50$ and $L = 80\pi$. The approximate spectral degree of coherence is illustrated in part (b) of Figure \ref{fig2} for $N = 80$, obtained from substituting \eqref{wapp} and \eqref{Sapp} into \eqref{defmu}. Clearly, as more coherence parameters are considered, the better the approximate correlation function is. Also, since the Fourier expansion is trying to match a two-dimensional truncated function (at $z_{1,2} \sim 0,L$ the function ends abruptly), it is expected that small oscillations appear due to the finite sum with low-frequency components. Parts (c) and (d) of Figure \ref{fig2} show the behavior of $\mu_{\text{app}}$ at the line $z_2 = L/2$ for $N = 10$ and $N = 80$. We verified that for all values of $N$, the condition $|\mu| \leq 1$ is satisfied. The distribution of the coherence parameters $d_{nm}$ in the $(n,m)$-plane is shown in Figure \ref{fig3} for the case illustrated in part (d) of Figure \ref{fig2}. From Figure \ref{fig3} it is seen that most of the $(2N+1)^2$ coherence parameters are nearly zero and only a small portion of coefficients around $n = m$ (with some oscillations) contributes to the sum. 

\begin{figure}[hbt]
    \centering
    \includegraphics[scale=0.25]{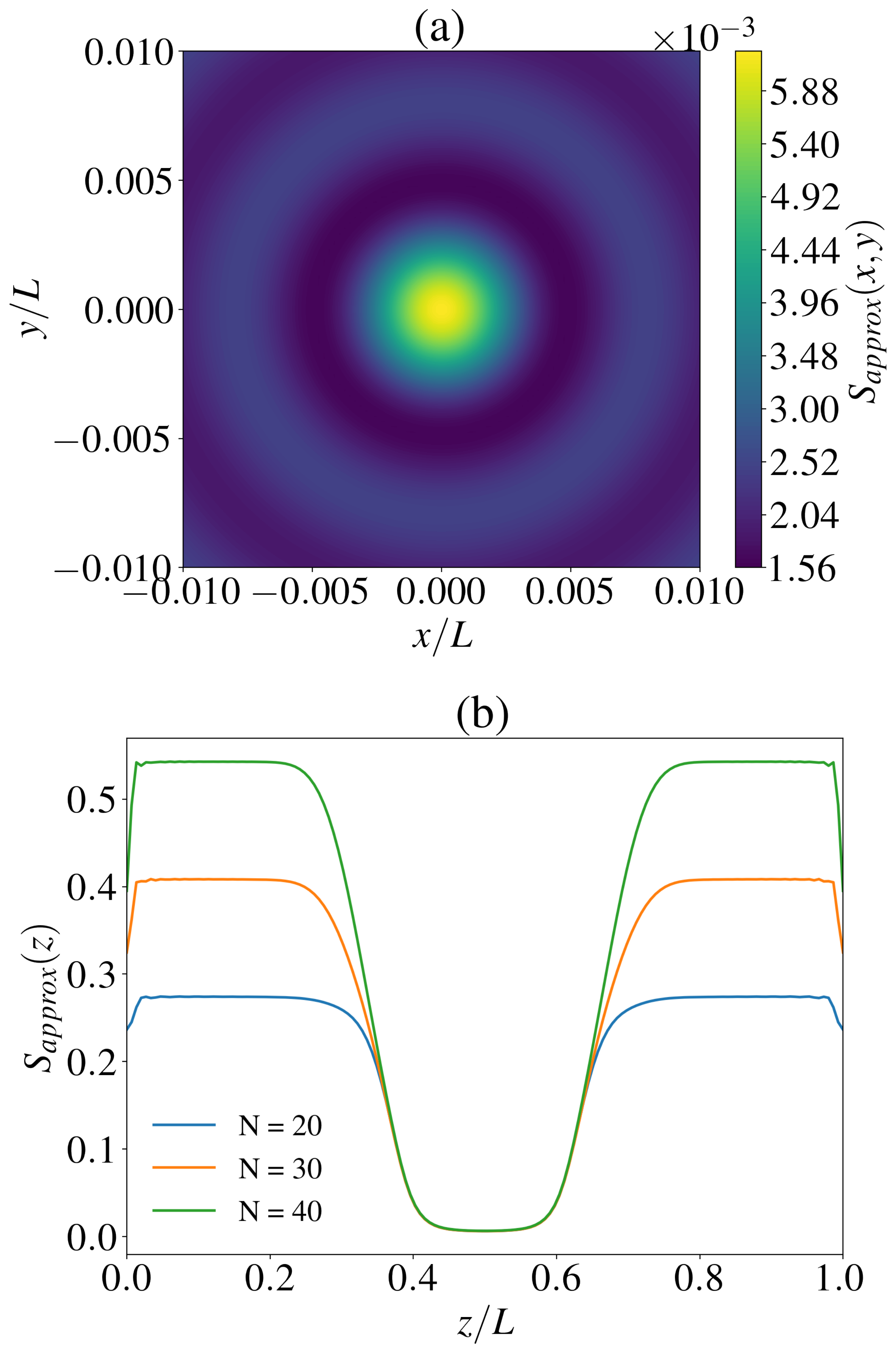}
    \caption{(a) Spectral density $S(x,y)$ at $z=L/2$ with $N=30$ and (b) spectral density on the propagation $z$ axis for three distinct values of $N$. Parameters: $\Delta = 50$, $L = 80\pi$, $\kappa=10^{-2}$ and $Q = 1/2 + 2\pi N / L$.} 
    \label{fig4}
\end{figure}

The intensity field distribution in a transverse plane is given by the spectral density $S_{\text{app}}(\mathbf{r})$ in \eqref{Sapp}. Part (a) of Figure \ref{fig4} shows the plot of the spectral density for the same coherence parameters $d_{nm}$ used in Figure \ref{fig3}. Due to the general form of \eqref{Sapp}, the resulting field undergoes self-imaging effects. The intensity distribution along the propagation axis, shown in part (b) of Figure \ref{fig4} displays a very curious behavior. Although we assumed $S(z)$ to be constant, it should be noted that it is not the spectral density function alone that is being represented as a Fourier series. The fact that we imposed $S(z) \sim S_0$ is not a guarantee for the function to behave in this way. In fact, the relationship between a general $S(z)$ together with $\mu$ is not trivial, regarding its Fourier series representation. Details on the internal structure of this more general model will be published elsewhere. The general behavior we found is that there is a dip with constant (non-zero) intensity around the region where the field is highly coherent, as shown in part (b) of Figure \ref{fig4}. For other values of $N$ the behavior is quite erratic involving non-trivial oscillations. We do not expect to provide a general explanation for these results in this article, focusing only on the possibility of creating a predetermined coherence profile.

A subtle and important point arises if one tries to reproduce discontinuous correlation functions. Suppose the objective is to approximate the correlation profile
\begin{equation}\label{gibbs}
    \mu(z_1,z_2) = 
    \begin{cases}
        1 &\text{if $z_1 = z_2$},\\ 
        1 &\text{if $(z_1,z_2) \in U$},\\
        0 &\text{otherwise},
    \end{cases}
\end{equation}
where $U$ is the set of points $(z_1,z_2)$ satisfying $(z_1 - L/2)^2 + (z_2 - L/2)^2 \leq R^2$, with $R$ positive. Even though there seems to be nothing suspicious about this structure, a Fourier decomposition of it to approximate the spectral degree of coherence is not possible as it violates the condition $|\mu_{\text{app}}(z_1,z_2)| \leq 1$ for pair of points near the discontinuity and also for some values of $N$ (not shown). Such behavior, known as Gibbs phenomenon, occurs in a general fashion for discontinuous functions and the approximate value always overshoot the true value of the discontinuity, even in the limit $N \rightarrow \infty$ (see Chapter 1, sec. 2 of \cite{sommerfeld1949partial}). The overshooting effect present in the Gibbs phenomenon is what precludes the use of models such as \eqref{gibbs} to represent spectral degrees of coherence in partially coherent frozen waves. This is the main reason behind our choice for the continuous model given by \eqref{mu1}. In principle, no such restrictions apply to the cross-spectral density $W(\mathbf{r}_1,\mathbf{r}_2)$.

In principle, the creation of a frozen spatial coherence should be easy to implement in a laboratory. There are many published schemes dealing with the generation and propagation of partially coherent Bessel beams \cite{liang2018high,xiao2019generation,santamaria2019generation}. We believe that the real challenge lies in verify that some specified region in the beam is really described by the predetermined coherence profile $\mu_{\text{app}}(z_1,z_2)$. The usual way to measure the degree of coherence between a pair of points is to match them into a double slit experiment format and look for the formation of an interference pattern. The visibility of the pattern generated by the light diffracted through the slits is thus directly related to the degree of coherence between $z_1$ and $z_2$ \cite{wolf2007introduction}. A discussion about the exact experimental implementation is not trivial and since it lies beyond the scope of the present paper, we leave it to the future.

In conclusion, we demonstrated that it is possible to formulate a partially coherent theory involving the superposition of partially coherent Bessel beams such that a desired spectral degree of coherence can be formed along the propagating axis. An explicit example in which an optical beam is highly coherent only if the pair of points $(z_1,z_2)$, along the axis, is inside a finite region around $z \sim L/2$, with $L$ being the propagation length. During the finalization process of our results, we became aware of a recent and interesting publication by F. Gori et. al., where the on-axis cross-spectral density was obtained for a particular effect involving self-focusing \cite{gori2022z}. Our results may implement and perhaps shed some new light regarding the engineering of these types of processes.

\begin{backmatter}
\bmsection{Funding} Content in the funding section will be generated entirely from details submitted to Prism. 
%\bmsection{Acknowledgments} The authors acknowledge the financial support of CNPq (Conselho Nacional de Desenvolvimento Científico e Tecnológico)

\bmsection{Disclosures} The authors declare no conflicts of interest.

\bmsection{Data availability} Data underlying the results presented in this paper are
not publicly available at this time but may be obtained from the authors upon
reasonable request.

%\bmsection{Supplemental document}
%See Supplement 1 for supporting content. 

\end{backmatter}

% Bibliography
\bibliography{sample}

% Full bibliography added automatically for Optics Letters submissions; the following line will simply be ignored if submitting to other journals.
% Note that this extra page will not count against page length
\bibliographyfullrefs{sample}

%Manual citation list
%\begin{thebibliography}{1}
%\bibitem{Zhang:14}
%Y.~Zhang, S.~Qiao, L.~Sun, Q.~W. Shi, W.~Huang, %L.~Li, and Z.~Yang,
 % \enquote{Photoinduced active terahertz metamaterials with nanostructured
  %vanadium dioxide film deposited by sol-gel method,} Opt. Express \textbf{22},
  %11070--11078 (2014).
%\end{thebibliography}

% Please include bios and photos of all authors for aop articles
\ifthenelse{\equal{\journalref}{aop}}{%
\section*{Author Biographies}
\begingroup
\setlength\intextsep{0pt}
\begin{minipage}[t][6.3cm][t]{1.0\textwidth} % Adjust height [6.3cm] as required for separation of bio photos.
  \begin{wrapfigure}{L}{0.25\textwidth}
    \includegraphics[width=0.25\textwidth]{john_smith.eps}
  \end{wrapfigure}
  \noindent
  {\bfseries John Smith} received his BSc (Mathematics) in 2000 from The University of Maryland. His research interests include lasers and optics.
\end{minipage}
\begin{minipage}{1.0\textwidth}
  \begin{wrapfigure}{L}{0.25\textwidth}
    \includegraphics[width=0.25\textwidth]{alice_smith.eps}
  \end{wrapfigure}
  \noindent
  {\bfseries Alice Smith} also received her BSc (Mathematics) in 2000 from The University of Maryland. Her research interests also include lasers and optics.
\end{minipage}
\endgroup
}{}

\end{document}